\documentclass[12pt,preprint]{aastex}

\newcommand{\logg}{\ensuremath{\log{g}}}

\newcommand{\rstar}{\ensuremath{R_\star}}

\newcommand{\rjup}{\ensuremath{R_{\rm J}}}

\newcommand{\teff}{\ensuremath{T_{\rm eff}}}
\newcommand{\rhostar}{\ensuremath{\bar{\rho_\star}}}

\slugcomment{Accepted to ApJ}
\shortauthors{Howell et al.}
\shorttitle{{\it Kepler} Exoplanet Candidates}

\begin{document}

\title{
{\it Kepler} Observations of Three Pre-Launch Exoplanet Candidates:
Discovery of Two Eclipsing Binaries and a New Exoplanet
}

\author{
Steve B. Howell \altaffilmark{1,8}
Jason F. Rowe \altaffilmark{2,9}
William Sherry \altaffilmark{3}
Kaspar von Braun \altaffilmark{4} \\
David R. Ciardi \altaffilmark{4}
Stephen T. Bryson \altaffilmark{2}
John J. Feldmeier \altaffilmark{5}
Elliott Horch \altaffilmark{6}
Gerard T. van Belle \altaffilmark{7}
}

\altaffiltext{1}{National Optical Astronomy Observatory, 950 N. Cherry Ave, Tucson, AZ 85719}
\altaffiltext{2}{NASA Ames Research Center, Moffett Field, CA 94035}
\altaffiltext{3}{National Solar Observatory, 950 N. Cherry Ave, Tucson, AZ 85719}
\altaffiltext{4}{NASA Exoplanet Science Institute,  California Institute of Technology,
770 South Wilson Avenue Pasadena, CA 91125}
\altaffiltext{5}{Department of Physics and Astronomy, Youngstown State University,
Youngstown, OH, 44555}
\altaffiltext{6}{Southern Connecticut State University, New Haven, CT 06515}
\altaffiltext{7}{European Southern Observatory, Karl-Schwarzschild-Str 2., 85478
Garching, Germany}
\altaffiltext{8}{Visiting Astronomer, Kitt Peak National Observatory, National Optical
Astronomy Observatory, which is operated by the Association of Universities for
Research in Astronomy (AURA) under cooperative agreement with the National Science
Foundation} 
\altaffiltext{9}{NASA Postdoctoral Program Fellow}


\begin{abstract} Three transiting exoplanet candidate stars were 
discovered in a ground-based photometric survey prior to the launch of
NASA's {\it Kepler} mission. {\it Kepler} observations of them were
obtained  during Quarter 1 of the {\it Kepler} mission. All three stars are
faint by  radial velocity follow-up standards, so we have examined these
candidates  with regard to eliminating false positives and providing high
confidence exoplanet selection. We present a first attempt to exclude false
positives for this set of faint stars without high resolution radial
velocity analysis. This method of exoplanet confirmation will form a large
part of the {\it Kepler} mission follow-up for Jupiter-sized exoplanet
candidates orbiting faint stars. Using the {\it Kepler} light curves and
pixel data, as well as medium resolution  reconnaissance spectroscopy and
speckle imaging, we find that two of our candidates are binary stars. One
consists of  a late-F star with an early M companion while the other is a
K0  star plus a late M-dwarf/brown dwarf in a 19-day elliptical orbit. The
third candidate (BOKS-1) is a $r$=15 G8V star hosting a newly discovered
exoplanet  with a radius of 1.12 R$_{Jupiter}$ in a 3.9 day orbit.
\end{abstract}

\section{Introduction}

Slightly more than two years before the launch of the NASA {\it Kepler}
mission (Borucki et al., 2010a), a 40 night  photometric variability study
was undertaken in the {\it Kepler} field of view (Howell 2008).  The
Burrell Schmidt Optical Kepler Survey (BOKS) consisted of  imaging a 1.25
square degree patch of sky centered on the open cluster NGC 6811 and was
aimed at the detection of variable stars and exoplanet transits. Data were
obtained using 3 minute SDSS r band exposures with a few V band exposures
to provide color information and relate  sources to other existing
catalogues of stars in the field. The BOKS survey magnitude range is r=14
to 20 with  over 35,000 point sources being measured and having light
curves produced.  Full details of the BOKS survey and our exoplanet transit
candidate detection procedures are presented in Feldmeier et al. (2010).

This paper describes {\it Kepler} observations of three exoplanet candidate
stars initially identified in the BOKS survey  as having light curves
suggestive of transit events. The three candidates had their discovery
light curves and initial vetting described in  Feldmeier et al. (2010).
Since that time, ground-based reconnaissance spectra and high resolution
imaging have been obtained and the sources were observed by {\it Kepler}
during Quarter 1 (Q1) operation. The transit events (initially discovered
in the ground-based survey) were confirmed in the {\it Kepler} data.  The
near-continuous observations and high photometric precision of {\it Kepler}
have enabled us to quantify the observed events and to understand the
candidate systems. While our analyses presented herein do not provide, by
themselves, complete  solutions for the non-exoplanet systems, they are
nevertheless easily characterized as false positives.  Additional
ground-based spectroscopy, in the future, could be used to fully describe
the non-exoplanet systems.

Our goal in this paper is to use the powerful constraints available from
the  analysis of the {\it Kepler} light curves and image data coupled with
high quality medium resolution spectroscopy and high resolution images to
eliminate false positives and provide high confidence identification. This
type of analysis and confirmation will be used in general for {\it Kepler}
exoplanet candidates that are large (Jupiter-like planets of which $>$100
are currently known; Borucki et al., 2010b) and have SDSS r magnitudes
fainter than $\sim$13.5. The large number of candidates and the relative
faintness of their host stars essentially eliminates the ability to obtain
enough large aperture high-resolution spectra  to use for planet mass
determination. Thus, the statistical value of {\it Kepler} results for
large planets orbiting faint stars will rely on high levels of false
positive elimination, and not on  radial velocity measurements of the
planet masses. 

False positive elimination begins with a number of steps involving  {\it
Kepler} data itself. Using pipeline produced light curves (Jenkins et al.
2010a),  and image data products (Bryson et al. 2010), candidates with at
least three consistent events have transit models fit to their phase
folding light curves.  Additionally, detailed  analysis of a number of
aspects of the light curves,  using difference imaging, and searching for
centroid  position shifts in the {\it Kepler} pixel data during transit are
also conducted. The details of the torturous false positive elimination
path the Kepler data follows prior to  ground-based observations is
discussed in Batalha et al. (2010) and  Borucki et al. (2010b). Only after
these steps have been passed are ground-based observations brought into
play, especially spectroscopy and high resolution  imaging. 

The three stars are referred to herein by  their {\it Kepler}
identification number (KID) as given in the Kepler Input Catalog  (KIC;
Batalha et al. 2010). We present finding charts for each object in Figure
1  and coordinates and photometric information for them in Table 1. From
our detailed analysis of the {\it Kepler} light curves and ancillary data
and additional  ground-based observations described below, we find that two
of the original exoplanet  candidates identified in the BOKS survey are
eclipsing binaries while the third is found to be a true exoplanet
system.  

Sections 2-4 deal with our observations using {\it Kepler}, ground-based
spectroscopy, and high resolution speckle imaging. Section 5 presents
conclusions for each candidate based on the observations and data analysis
while Section 6 concludes the paper.

\section{{\it Kepler} Photometric Observations}

The {\it Kepler} mission and its photometric performance since launch are
described in  Borucki et al. (2010) while the CCD imager on-board {\it
Kepler} is described in  Koch et al. (2010) and Van Cleve (2008). The
observations used herein consist of  data covering a time period of 32
days, obtained during {\it Kepler's} Quarter 1 of operation (May-June 2009
corresponding to HJD 2454964 to 2454998). The {\it Kepler} photometry of
our three stars are presented in Figures 2-4.  The photometric observations
were reduced using the {\it Kepler} mission data pipeline  (Jenkins et al.,
2010) and then passed through various consistency checks and exoplanet
transit detection software as described in Van Cleve (2009). Normalized and
phase-folded light curves were then produced for each of the three BOKS
exoplanet candidates.  The transit events in the phased light curves were
modeled in an effort to understand the transiting companion, both from the
primary transit event and the secondary eclipse, if observed. Two of the
candidates (KIC~9838975 and KIC~9597095) show clear evidence for a
secondary eclipse and  are consistent with being eclipsing binaries. The
remaining candidate (KIC~9838975)  is revealed to be a bona fide
exoplanet. 

\subsection{Transit Fits}

The {\it Kepler} photometric time series are initially fit to a transit
(when the companion moves in front of the star) model to determine the
scaled radius ($R_p$/\rstar), scaled semi-major axis (a/\rstar), impact
parameter (b) and the period (P) and epoch (T0) of the companion.  Our
transit model uses the analytic formulae of Mandel \& Agol (2002) for a
non-linear limb-darkening law and best fits are found using a
Levenberg-Marquardt algorithm (Press et al., 1992). Limb-darkening
coefficients where adopted from fits to Atlas 9 spectra (Sbordone et al.
2004) convolved with the Kepler bandpass (Prsa 2010). We model the
occultation (when the companion moves behind the star) by computing the
fraction of the companion occulted by the star as a function of the
star-planet projected distance and fitting for the brightness of the
companion relative to the star.  We assume the planet is a uniformly
illuminated disc during occultation.  In cases where we detect a
significant occultation we allow for eccentric orbits and appropriately
adjust the depth of the transit model to account for dilution. 
Uncertainties for the model parameters were derived from a
Markov-Chain-Monte-Carlo (MCMC) analysis (Ford 2005). We refer the reader
to Koch et al. (2010) for a full description of the transit modeling
procedure.

We also compute an odd-even transit depth and occultation depth statistic
that provide useful diagnostics in the identification of stellar binaries. 
The odd-even statistic determines the significance of a change of depth of
odd and even numbered transits.  If the transit depths are found to be
changing in such a fashion, then we have strong evidence that we are seeing
a stellar binary where the two components have slightly different surface
flux densities and the true orbital period is twice as long.  The odd-even
effects where measured at 0.7, 1.6 and 0.5 sigma confidence levels for
KIC~9595827, KIC~9838975 and KIC~9597095 respectively.  No significant
odd-even effect was found for the three candidates described in this paper.

The occultation depth statistic attempts to measure the significance of an
occultation of the companion.  The search is done by first removing
photometric measurements during transit and then phasing the remaining data
at the orbital period and computing the maximum displacement of a segment
of phased data with a length equal to the transit duration.  No
displacements are found for KIC~9595827 (the largest displacement had a 0.7
sigma significance) level while strong displacements were identified for
KIC~9597095 and KIC~9838975.

The scaled semi-major axis is related to the mean density of the
primary star and companion by,
\begin{equation}\label{eq:rhostar}
\left(\frac{a}{R_\star}\right)^3 \frac{\pi}{3 G P^2} =
\frac{(M_\star+M_p)}{\frac{4 \pi}{3}R_\star^3}.
\end{equation}
If $M_\star >> M_p$, then it is a simple matter to estimate the mean
stellar density (\rhostar) of the primary star and deduce its stellar
parameters by matching to a set of theoretical stellar evolution models
with an independent measurement of the stars effective temperature
(Sozzetti et al. 2007).  For KIC~9838975 and KIC~9597095, the depth of the
occultation is used to estimate the temperature of the companion, by
assuming the companion radiates as a blackbody and correcting for the
Kepler bandpass.  We estimate temperatures for the companion of $\sim$ 2700
K and $\sim$ 4000 K respectively.  Such temperatures suggest companions is
with masses of $\sim$0.1 $M_\sun$ and $\sim$0.5 $M_\sun$.  For a
Jupiter-mass companion an error of 0.02\% will be incurred on the
measurement of \rhostar, a $0.1 M_\sun$ companion would skew our estimate
of \rhostar\ by $\sim$ 2\% and a 0.5 $M_\sun$ companion would induce a
systematic error of 41\% on \rhostar.  It should also be noted that we have
assumed a circular orbit for KIC~9595827.  Should the companion exist in an
eccentric orbit, our estimate of the scaled semi-major axis could exhibit
large systematic errors (c.f., Tingley, Bonomo and Deeg, 2010).

We adopt of estimates for the masses of the companions listed above and
calculate \rhostar.  For KIC~9597095 we also carry through an error of 10\%
to account for our uncertainty in the mass of the companion.  We then
proceed to determine the stellar parameters of the primary stars by
matching \teff\ (see Table 2) with an adopted error of 150 K with the mean
stellar density to the Yonsei-Yale stellar evolution tracks (Demarque 2004)
to calculate the range of stellar parameters consistent with the
observation (e.g. Borucki et al. 2010b).  This calculation produces a
Markov-Chain of 100 000 samples of $M_\star$, $R_\star$ pairs.  These
values are then used to determine new fits for T0, P, $R_p$, orbital
inclination (i) and occultation depth as well as uncertainties in our model
fits to light curves as listed in Table 3.

\subsection{Image Analysis}

Examination of the Kepler pixel images within the aperture of an exoplanet
candidate star provides powerful extra constraints that can eliminate false
positives. The stars KIC~9597095 and KIC~9838975 have already been
eliminated as false positives for exoplanet candidates by virtue of their
{\it Kepler} light curves alone. However, the exoplanet candidate
KIC~9595827, {\it not shown}  to be a false positive event based on the
folded light curves and transit models discussed above, can be subjected to
additional testing by making use of the downloaded image  pixel data. The
use of {\it Kepler} image pixel data for false positive elimination is
described in Batalha et al. (2010). Details of initial image centroiding
and difference image use are discussed  by Jenkins et al. (2010b) and make
for additional methods by which false positives can be identified. We have
improved on those techniques as discussed below.

The pixel image aperture containing KIC~9595827 is shown in Figure 5 (top
two panels) and is seen to contain 24 pixels in a 6X4 arrangement. 
Formation of these two images is done by first detrending then folding the
pixel time series and finally averaging over several in-transit (5 in this
case) and  out-of-transit (20) points. This improvement from the original
analysis provides more robust images. The small change in flux of the
target star between in and out of transit is too small to be noted here
given the image stretch and the fact that a {\it Kepler} star is purposely
slightly out of focus with a FWHM value near 4 arcsec (see Bryson et al.
2010). Since each pixel is 3.98 arcsec in size (Koch et al. 2010), the
total aperture used here is 24 by 16 arcsec and is seen to contain our
target star and a piece of a close companion. This brighter neighbor  star
can also be seen in Figure 1. The central source seen in Figure 5 is our
target, KIC~9595827, while the other source, only partially in the
aperture  (at 1090, 552) is KIC~9595844 with r$\sim$13. This star has its
own {\it Kepler} light curve and shows no variation consistent with the
transit observed in KIC~9595827.  Its constant third light contribution to
the total aperture flux is responsible for, and consistent with, the
observed  very small centroid shift seen in KIC~9595827 during transit
(-0.43$\pm$0.06, -0.99$\pm$0.09 millipixels). 

Making use of {\it Kepler} pixel images for KIC~9595827  when in and out of
transit, we can form a difference image and check to see if the target star
is indeed the one changing in flux during the transit.  We note in Figure 1
an additional close, faint companion to KIC~9595827 that will be well
blended with the target star in the Kepler aperture.  This star is 2.5-3
magnitudes fainter than the primary object and would, if an eclipsing
binary,  have to have an eclipse depth of 14\% in order to mimic the
observed transit.  This change in light would easily be observed in the
difference image but is not seen. The bottom left panel of Figure 5 shows
the difference image and reveals that indeed only KIC~9595827 varies during
the transit.

A final and very informative use we can make from the pixel data is to
produce individual pixel time series light curves. These are shown in
Figure 6 in the same pixel arrangement as that of the {\it Kepler} pixel
images in Figure 5. The transit event is clearly see in the pixels
containing KIC~9595827 but not observed in the brighter companion star
partially in the aperture.

\section{Spectroscopic Observations}

Optical spectroscopy for our three candidate exoplanet stars was obtained
using the Kitt Peak 2.1-m telescope and GCAM spectrograph and using the
Kitt Peak 4-m telescope and RCSpec instrument. Spectra were obtained in
June 2008 on both telescopes as well as in May 2010 on the 4-m to search
for evidence of binarity and any spectral changes due to eclipsing binary
motion, lines shapes, and/or RV variations. The 2-m GCAM setup used a 300
l/mm grating (\#32) with a 1 arcsec slit to provide a mean spectral
resolution of 2.4\AA~per resolution element  across the full wavelength
range. Both 4-m RCSpec setups used a 632 l/mm grating (KPC-22b in second
order) with a 1 arcsec slit to provide a mean spectral resolution of
1.6\AA~per resolution element across the full wavelength range. The spectra
were reduced in the normal manner with observations of calibration lamps
and spectrophotometric stars (obtained before and after each sequence) and
bias and flat frames collected each afternoon. 

The resulting spectra were compared with MK standard stars digitally
available in the ``Jacoby Atlas" (Jacoby et al. 1984) and spectral types
and luminosity classes were estimated by comparison to all stars in the
atlas (O to M and of luminosity classes I, III, IV, and V) via  a standard
$\chi^2$ fitting technique.  Gaussian line fits to the strong  spectral
features were used to determine line centers, providing a velocity
resolution of approximately one-tenth of a pixel across the full wavelength
range in the 4-m spectra, or about 7 km/sec. Example spectra for each
object are shown in Figure 7. Table 2 presents our spectral type and
effective temperature estimates based on the Kitt Peak observations, as
well as other KIC derived intrinsic properties for the candidate stars.

\section{Speckle Observations}

One of the most probable and devious false positive scenarios for
photometric transit exoplanet  surveys is the presence of a nearby
companion star, either a real companion or a  line-of-sight eclipsing
binary. For a Jupiter depth transit event ($\sim$1\%), a companion leading
to a false positive must be within a few magnitudes of the target star as
well as within about one arcsecond. If a background star is fainter or
further away, it cannot produce such a deep transit. The {\it Kepler}
photometry alone eliminates much of the phase space here via centroiding
measures of the pixel data in and out of the transit event. Typical direct
imaging can identify companions as close as about 1-1.5 arcsec and to
perhaps 6-8 magnitudes fainter. Medium resolution spectroscopy can identify
both orbital motion due to a stellar companion or direct evidence for two
stars via a composite spectrum.  However, for companions closer than 1
arcsecond and to about 5-6 magnitudes fainter, optical speckle observations
are generally required to fully eliminate possible confounding sources. 
Techniques as deblending the images and fitting triple star scenarios are
very powerful as well and have been successfully applied to exoplanet
transit systems (DECPHOT (Weldrake et al. 2008); BLENDER (Torres et al.
2010)).

Speckle observations form a major part of the {\it Kepler} mission false
positive elimination strategy and are fully described in Horch, et al.,
(2009) and Howell et al. (2010).  We make use here of the Differential
Speckle Survey Instrument, a recently upgraded speckle camera described in
Horch et al. (2010) that provides simultaneous observations in two filters.
The speckle instrument uses two identical EMCCDs and generally observes in
"V" and "R" bandpasses where "V"  has a central wavelength of 5620\AA~and a
FWHM=400\AA~and "R" has a central wavelength of 6920\AA~and a FWHM=400\AA.
Our speckle observations of KID~9595827 reported here  were obtained on 21
\& 22 June 2010 UT with the WIYN 3.5-m telescope located on Kitt Peak. On
both nights we obtained 8 sets of 1000 simultaneous "V" and "R" images
each  using 40 msec and 80 msec  frame times respectively. KID~9595827, at
Kepmag=15.1, is a fairly faint and challenging speckle target as well as
being far too faint for high-resolution spectroscopic observations. The
sets of speckle images were, as usual, co-added in Fourier space and
re-projected into real space to produce the final reconstructed images (see
Tokovinin et al. 2010).

Figure 8 shows the final reconstructed "R" speckle image of KID~9595827
from 22 June 2010 UT. The two images at the bottom of the plot are
identical but presented with  different image stretch values for ease of
use and visual identification of any possible companions.  The "cross"
pattern in each image is an artifact of the reconstruction process.

\section{Discussion}

The three {\it Kepler} light curves presented in Figures 2-4 and their
model fits tell us most of the story. The additional ground-based
observations provide confirmation and some additional constraints for our
three systems but the {\it Kepler} light curves alone get us about 90\% of
the way to a robust solution. The parameters from our model fits are listed
in Table 3.

Table 1 provides KIC catalog information for the three stars\footnote{Note
that the KIC is available at the MAST archives
(http://archive.stsci.edu/)}. Table 2 has the derived stellar parameters
for the three stars based on the catalog KIC colors and model fitting with
the spectral type being determined from our spectra. The observed spectral
type are in very good agreement with the temperatures determined in the
KIC. Table 3 summarizes the orbital elements for the companions. Since our
ground-based survey field was centered near the open cluster NGC 6811, we
were interested to see if any of the stars under study here were likely
cluster members. None of the stars are  at the cluster distance ($\sim$1200
pc) and, in addition, they all lie well outside the cluster core radius.
Thus, we do not believe that these three stars  are cluster members.

\subsection{KID-9597095}\label{EB}

The BOKS survey  discovered this source as a r=16 variable showing a few
percent  exoplanet like transit with a period near 2.7 days. The star,
designated BOKS-r.45069, is revealed through optical spectroscopy (Figure
7) to be a F8-F9 main sequence star with an effective temperature near
6400K and a distance estimate of  2500 pc. There is no evidence in our
spectrum of a secondary star suggesting a large luminosity ratio. The {\it
Kepler} light curve (Fig. 2) shows an easily noted V shaped  primary
eclipse and a clear secondary eclipse providing a deterministic diagnostic
that  this star is an grazing eclipsing binary. Knowing the primary star
parameters, the eclipses allow the companion star and its orbit to be
modeled. The secondary star has a radius of 5.9 R$_{Jupiter}$ with an
effective temperature near  4100K, consistent with a late K or early
M-dwarf, having 0.83\% of the luminosity of  the F star primary.  The
orbital inclination is 79 degrees and the companion is likely heated by its
close proximity to the F star. We note that the secondary eclipse depth in
this system is  $\sim$1\% a level which is challenging but doable from the
ground (c.f., Weldrake et al., 2008).  The low mass companion star will be
greatly out-shined by the F star in the optical and near-IR and a radial
velocity solution for the binary will be required to fully determine the
secondary star mass.

There is variability related to the orbital period of the companion that
has a peak-to-peak amplitude of  $\sim$1.8 mmag.  Measuring the first two
harmonics at 2.75 d and 1.37 d gives amplitudes of 0.7 and 1.8 mmag
respectively.  The second harmonic shows amplitude modulation over the
timescale of the {\it Kepler} Q1 observations.  If we interpret the cause
of these interactions as ellipsoid distortions and Doppler boosting (Rowe
et al. 2010 \& van Kerkwijk et al. 2010) then we can estimate a mass of
approximately 0.4 $M_\sun$.

\subsection{KID-9838975}\label{koi218}

Discovered in the BOKS survey, this r=16 star  showed an exoplanet
transit-like event in its light curve but with only one transit observed,
its orbital period was unknown. Spectroscopy (Figure 7) determines the
star, designated as BOKS-r.45069,  to be a K0V($\pm$one subclass in
spectral type) with an effective temperature near 5000K and a distance of
$\sim$1 kpc.  The {\it Kepler} light curve (Fig. 3) shows a U-shaped
transit event as well as a clearly off centered secondary eclipse
suggestive of an elliptical orbit. The secondary eclipse in this system
would be extremely difficult to observe from the ground as it has a depth 
near one-half of one  percent. Thus, this object would be extremely
difficult to eliminate as an exoplanet using ground-based photometry alone.
The companion causing the transit event has a radius of 1.4 R$_{Jupiter}$
and a modeled effective temperature of 2760K. The orbital eccentricity is
$\sim$0.1, making the difference between aphelion and perihelion change by
approximately 20\%.   Thus, at an average distance from the K0V star of
$\sim$0.13 A.U., the companion is not close enough for irradiation to
account for the observed temperature. At this low effective temperature, a
low mass red or brown dwarf  is the most likely companion star in a near 90
degree elliptical orbit with a period of 18.7 days.  Additional {\it
Kepler} observations and a radial velocity study should be able to  easily
confirm the elliptical nature of the orbit as well as provide a good mass
estimate for the companion star.

\subsection{KID-9595827 (BOKS-1)}\label{koi217}

The best exoplanet candidate from the BOKS survey, BOKS-1, is a r=15
variable showing an exoplanet transit-like event  with a period near 4
days. Ground-based spectroscopy (Figure 7) shows the host star to be a
G8V($\pm$one subclass) with an effective temperature of 5500K at a distance
of $\sim$800 pc.  Spectroscopic observations of this target obtained on two
consecutive nights in June 2008 (roughly 0.5 apart in phase) and three
nights in May 2010,  reveal no statistically significant radial velocity
trends at the $\sim$7 km/sec level or larger. This limit rules out a
companion star with a mass greater than 0.1M$_{\odot}$ i.e., virtually all
dwarf stars. 

Our speckle imaging was aimed at a search for background (or real)
companions which may be the cause of the transit signal. To robustly
estimate the background limit we reach in each reconstructed speckle image,
we use the metrics derived from the top plot in Figure 8.  If a companion
is present, it will appear in the reconstructed image as a "point" like
source, similar to the central source. As in a normal image, the value of
pixels in the target and any companion star images will lie above the local
"sky" background. If a nearby source appears and is  approximately 3-5
sigma above the background (the line in the top plot is a 5 sigma limit
from the mean of the distribution of background points) we will easily see
it and can robustly measure its properties. If, however, a source is barely
detected (say at 1.5 sigma), then we will just be able to note its
presence,  but would derive only approximate information for it. Local
minima are also plotted mainly as a check to see if the minimum and maximum
points are normally distributed (as would be the case for a well produced
reconstructed "sky"). If there were too many minima or they were too large
a value (+ or -) that could indicate a problem in the reconstruction.

Using our simultaneous 2-color speckle camera we can detect companions, for
bright targets, below the diffraction limit (0.035 arc second at 5000\AA)
by detection of a single fringe in both cameras that provides the same
solution. For fainter targets  (R$\ga$13.5-14.5) and good observing
conditions, our inner detection  limit ranges from 0.05-0.15" depending on
the magnitude difference of the companion. Thus, for multi-fringe images, a
very conservative inner limit for robust detections is near 0.2 arc
seconds. Our high resolution ground-based speckle observations allow us to
eliminate any possible companion stars to 4.2 magnitudes in R and 5.0
magnitudes in V fainter and from  i$\sim$0.05 to 1.8 arcsecond of
KID-9595827.  Thus the observed transit depth of 1.2\%  can not be mimicked
by a variable or eclipsing binary background star even if it showed a 50\%
dimming. Taken together with the {\it Kepler} light curve model results, no
companion star is present that can be the cause of the observed exoplanet
transit event. 

Image analysis of the {\it Kepler} pixel data for KIC~9595827 reveals that
it is the source of variation within its aperture during the transit event
and that its small centroid motion is consistent with the star itself
dimming in the presence of a near by constant light neighbor. Additionally,
pixel time series light curves clearly reveal that the transit event in
centered on our target star.  No evidence from the image data suggests any
explanation for the observed variation except as that of an exoplanet
transit. 

The {\it Kepler} light curve observations (Figure 4) show the typical
U-shaped exoplanet transit and reveal no evidence for a secondary eclipse.
The transiting body has a radius of 1.12 R$_{Jupiter}$ orbiting every 3.9
days at nearly 90 degrees to the plane of the sky. If this exoplanet is
similar to other ``hot Jupiters", it is likely to  have a mass near one to
a few Jupiter masses. The raw {\it Kepler} light curve shows evidence for a
rotational modulation  of the host  star with a period near 17-18 days and
we note possible evidence of star-spots  seen as slight wiggles at the
bottom of the exoplanet transit.

We tested the possibility of hierarchical triples, by adding a dilution
factor to our models.  The dilution factor accounts for inclusion of third
light from unseen stellar companion in the system. The effect is that the
observed transit can be much shallowed than reality.  For example, if two
stars of equal brightness are found within the Kepler aperture, the
measured transit would need to be made twice as deep.  We modeled the
transit of KID~9595827 with dilution factors ranging from 0 to 0.7, in
increments of 0.05, where 0.7 means 70\% of the observed flux comes from an
additional star within the photometric aperture. Comparing chi-squared
values we find models with dilution factors greater than 0.1 are ruled out
at the 99.97\% confidence level.  In particular the egress and ingress of
the models are incompatible with the size of the companion inferred from
the depth of the transit.  While, there may be third light contamination
due to a hierarchical triple scenario, the unseen companion must contribute
less that 10\% of the light in the photometric aperture.  At such dilution
factors the planet radius would increase by 5\% but still be compatible
with being a bone-fide planet.

In addition, the {\it Kepler} pixel data are not consistent with a nearly
perfectly aligned (within $\sim$0.05 arcsec) background eclipsing binary
(BGEB) being the cause of the transit.  To estimate the possibility of a
chance alignment, we can take the mean local star density observed near
KIC~9595827 (0.0122 stars/arcsec$^2$  to r$\sim$20)  times the likelihood
that a background star would be an eclipsing binary  (1.2\% of stars are
eclipsing binaries; Torres et al.  2010). The product of these values
yields a probability of only 0.015\% that a BGEB could be  the cause of the
transit event.  If we assume a near circular orbit for the stars in any
BGEB,  we can estimate this probability another way. The observed  transit
time restricts the background population of confounding EBs to only G8
stars or earlier.  Using the Besancon Galactic model  we calculate a
stellar density of possible (G8 and earlier) pollutants  to be 2e-4
stars/arcsec$^2$. Taking the same likelihood as above that a background
star could be an EB,  this approach suggests only a 0.00024\% chance that
the transit could be caused by a BGEB. Given that these conservative
estimates of a chance alignment of a BGEB or a hierarchical triple  can be
the cause of the observed transit event are essentially zero, we again
conclude that KIC~9595827 (BOKS-1) does indeed harbor an exoplanet.

%Furthermore, any BGEB capable of being undetected in our observational data yet
%producing the observed transit signal would need to be at a far distance
%($\sim$700 or more pc) and would suffer greatly from Galactic extinction 
%minimizing its flux contribution to our target star.

{\it Kepler} is still monitoring this source and revealing continuous 
transits leaving no doubt as to its periodic nature  and constant transit
shape and depth. Phase folding of all the data to date reveals no secondary
eclipse  eliminating essentially all possible binary star configurations.  
Future {\it Kepler} observations will hopefully detect the occultation
allowing one to determine the companions orbit. At r=15, it is unlikely
that high precision radial velocities will soon be forthcoming for this
system.

\section{Conclusions}

{\it Kepler} observations have provided unprecedented light curves for 
three pre-launch exoplanet candidate stars. Two of them, KID-9838975 and
KID-9597095 are shown to be main sequence plus late-type dwarf/brown dwarf
binaries, one harboring its companion  in a 19-day elliptical orbit. We
note that {\it Kepler} photometry was likely necessary in order to observe
the  secondary eclipse in KID9838975 and may ultimately detect the
occultation in KID9595827 (BOKS-1). Such small dips would unlikely be
distinguishable from the ground due to the necessary precision required
over their duration and covering a number of occultations. This type of
binary star light curve  suggests a word of caution for ground-based
exoplanet candidates, even those  with well determine light curves, as very
small secondary eclipses or occultations would easily be missed. The two
binary stars discovered here are ripe for radial velocity follow-up  in
order to determine a better mass  and orbital characteristics for the low
mass companions.

We have presented a first attempt to fully utilize {\it Kepler} light
curves and  {\it Kepler} image analysis combined with ground-based
follow-up observations to exclude false positive exoplanet candidates
without high-resolution spectroscopic radial velocity measurements. The
best pre-launch BOKS survey exoplanet transit candidate (BOKS-1) has been 
confirmed by this methodology to be  a true exoplanet; a ``hot jupiter"
having a radius of  1.12 R$_{Jupiter}$  and an orbital period near 3.9
days.   The {\it Kepler} light curves and pixel data analysis alone can
provide a very robust method to eliminate false positives and confirm, at a
high confidence level, Jupiter-size planets. However, ground-based
follow-up observations provide additional information and confirmation of
the {\it Kepler} results as well as placing  hard constraints on the
possible companions.  This system at r=15, will likely need to await the
next generation of larger telescope and/or a more sensitive radial velocity
instrument before its  exoplanet mass can be precisely determined.

\acknowledgments
We wish to thank Brandon Tingley for his timely review and very helpful
comments on  our original manuscript. The authors would like to thank the
{\it Kepler} Science Office and the Science Operations Center personal,
particularly Natalie Batalha Jon Jenkins, and Tim Brown, for their
dedicated effort to the mission and  for providing us access to the science
office data products. The ground-based observations reported on herein were
obtained at  Kitt Peak National Observatory, National Optical Astronomy
Observatory, which is operated by the Association of Universities for
Research in Astronomy (AURA) under cooperative agreement with the National
Science Foundation. The Kepler Science Team is thanked for their help and
support of the mission and its scientific output.  {\it Kepler} was
selected as the 10th mission of the Discovery Program.   Funding for this
mission is provided by NASA.

\newpage

%Table 1
\begin{deluxetable}{cccccccccc}
\rotate
\tablenum{1}
\tablecolumns{11}
\tablewidth{0in}
\tablecaption{Kepler Input Catalogue Photometry}
\tablehead{
\colhead{KepID} & \colhead{RA} & \colhead{Dec} & 
\colhead{r} & \colhead{g} & \colhead{i} & \colhead{z} &
\colhead{J} & \colhead{H} & \colhead{Ks}\\

& J2000 & J2000 & [mag] & [mag] & [mag] & [mag] & [mag] & [mag] & [mag] 
 }
 
\startdata
9597095 & 19 41 18.86 & +46 16 06.2 & 15.90 &  16.53 & 15.69 & \nodata & 14.91$\pm$0.04 &
14.69$\pm$0.06 & 14.69$\pm$0.11 \\
9838975 & 19 40 08.04 & +46 36 00.5 & 16.10 &  16.83 & 15.86 & 15.69 & 14.65$\pm$0.03 &
14.17$\pm$0.04 & 13.96$\pm$0.05 \\
9595827 & 19 39 27.72 & +46 17 09.1 & 15.06 &  15.69 & 14.89 & 14.80 & 13.93$\pm$0.03 &
13.55$\pm$0.02 & 13.47$\pm$0.04 \\
\enddata

\tablecomments{Magnitude uncertainties are $\sim$0.02 mag for r, i, g and $\sim$0.03 mag for
z}

\end{deluxetable}

%Table 2
\begin{deluxetable}{ccccc}
\tablenum{2}
\tablecolumns{5}
\tablewidth{0in}
\tablecaption{Summary of Stellar Parameters}
\tablehead{
\colhead{KepID} & \colhead{Temperature} & \colhead{Log(g)} & \colhead{Metallicity} &
\colhead{Spectral}\\
& [K] & [cm/s$^2$] & [solar=0] & Type
 }
 
\startdata
9597095 & 6400 & \nodata & \nodata & F8V-F9V \\
9838975 & 4967 & 4.802 & -0.511 & K0V \\
9595827 & 5545 & 4.724 & 0.22 & G8V \\
\enddata

\tablecomments{Temperature, log(g), and metallicity derived photometrically from KIC data.  
Spectral Type (and temperature for 9597095) 
derived from Kitt Peak spectra.}

\end{deluxetable}

%New Table 3
\begin{deluxetable}{lccc}
\tablenum{3}
%\tablecolumns{9}
\tablewidth{0in}
\tablecaption{Transit Model Derived Parameters}
\tablehead{
\colhead{ } & \colhead{9597095} & \colhead{9838975} & \colhead{9595827}
}
\startdata
Period (d) & 2.74560$\pm$0.00003 & 18.6929$\pm$0.00005 & 3.90512$\pm$0.00005\\
T$_0$ (HJD-2454900) & 64.7080$\pm$0.0002 & 76.8324$\pm$0.0004 & 66.4140$\pm$0.0002\\
M$_\star$ (M$_\odot$) & $1.21\pm0.03$ & $0.79^{+0.03}_{-0.05}$ & $0.95^{+0.04}_{-0.05}$\\
R$_\star$ (R$_\odot$) & $1.43\pm0.01$ & $0.72\pm 0.01$ & $0.86\pm 0.02$\\
R$_p$ (\rjup) & $5.9\pm 0.5$ & $1.43^{+0.02}_{-0.04}$ & $1.11\pm 0.02$\\
i (deg) & $79.4\pm 0.5$ & $89.7^{+0.2}_{-0.3}$ & $89.8^{+0.2}_{-0.4}$\\
e & 0 (fixed) & $0.15\pm0.02$ & 0 (fixed)\\
a (AU) & $0.0411\pm0.0003$ & $0.128\pm0.002$ & $0.0477\pm0.0007$ \\
$R_p/R_\star$ & $0.41\pm0.04$ & $0.2026^{+0.0006}_{-0.0011}$ & $0.1325^{+0.0003}_{-0.0005}$\\
\logg$_\star$ & $4.21\pm0.01$ & $4.61\pm0.01$ & $4.54\pm0.01$ \\
Secondary Depth (ppm)& $21033\pm3084$\tablenotemark{a}  & $1173\pm 228$ & $-33\pm 65$\\
\teff$_p$ (K) & $4050\pm 355$ & $2760\pm 186$ & $< 2200$ \\
\enddata
\tablenotetext{a}{Estimated depth if companion is fully occulted.}
\end{deluxetable}

%Table 3
%\begin{deluxetable}{cccccccc}
%\tablenum{3}
%\rotate
%\tablecolumns{9}
%\tablewidth{0in}
%\tablecaption{Transit Model Derived Parameters}
%\tablehead{
%\colhead{KepID} & \colhead{Period} & \colhead{T$_0$\tablenotemark{a}} & 
%\colhead{M$_\star$} & \colhead{R$_\star$} & \colhead{R$_\star$} & \colhead{R$_p$} &
%\colhead{T$_{{\rm eff}}$}\\
%
%& [Days] & [HJD] & [M$_\odot$] & [R$_\odot$] & [\rjup] & [R$_{Jup}$] & [K]
% }
% 
%\startdata
%9597095 & 2.74561$\pm$0.00003 & 64.7080$\pm$0.0002 & 5.0$\pm$0.0 &   &  & 2.62$\pm$0.01 &
%4139$\pm$318 \\
%9838975 & 18.6931$\pm$0.00005 & 76.8210$\pm$0.0009 & 5.0$\pm$0.0 &   &  & 1.39$\pm$0.01 &
%2090$\pm$186 \\
%9595827 & 3.90512$\pm$0.00005 & 66.4136$\pm$0.0002 & $0.95^{+0.06}_{-0.03}$ & $0.86\pm 0.01$ & $1.11\pm 0.02$& 1.12$\pm$0.01 &
%\nodata \\
%\enddata
%\tablenotetext{a}{T$_0$ = HJD - 2454900}
%
%
%\end{deluxetable}

\clearpage

%Fig. 1
\begin{figure}
%\plotone{K_find.ps}
\plotone{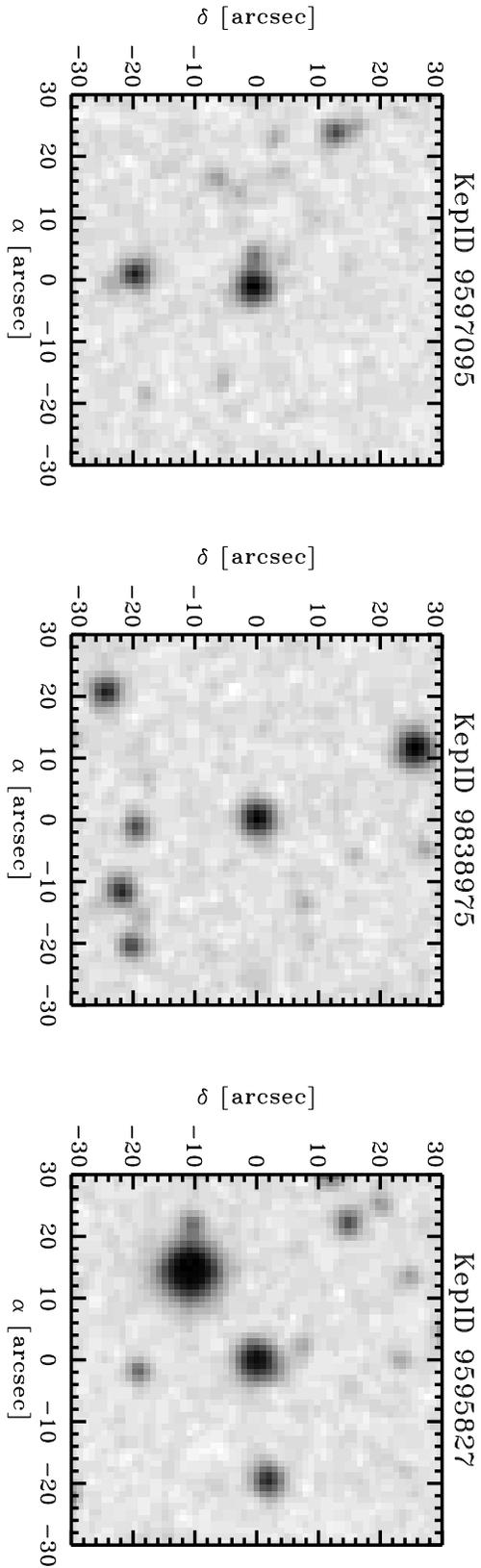}
\caption{Finder charts for the three BOKS stars discussed in this paper.
These charts show POSS2 Red images produced from the Digitized Sky Survey.}
\end{figure}

%Fig. 2
\begin{figure}
%\plotone{K_EB.eps}
\plotone{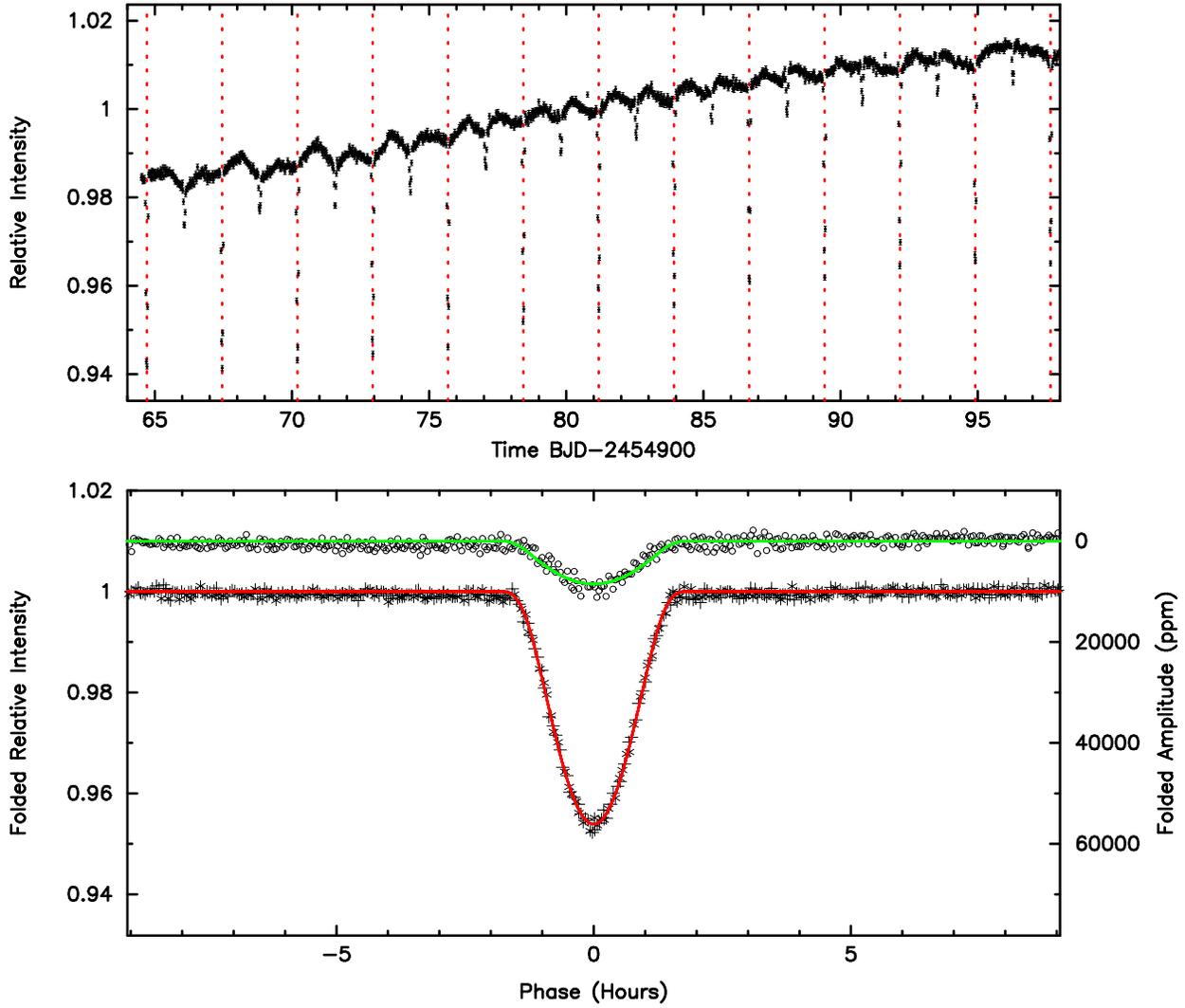}
\caption{{\it Kepler} light curve for KID-9597095. The top panel shows the normalized 
{\it Kepler} light curve while the bottom panel shows this light curve phase folded near transit
(Phase 0.0 - crosses) and near phase 0.5 (circles). The plot has two vertical axes scales, the left one is for the transit depth (black
crosses) while the right scale is for the secondary eclipse (grey circles). This object is an eclipsing binary.}
\end{figure}

%Fig. 3
\begin{figure}
%\plotone{K_218.ps}
\plotone{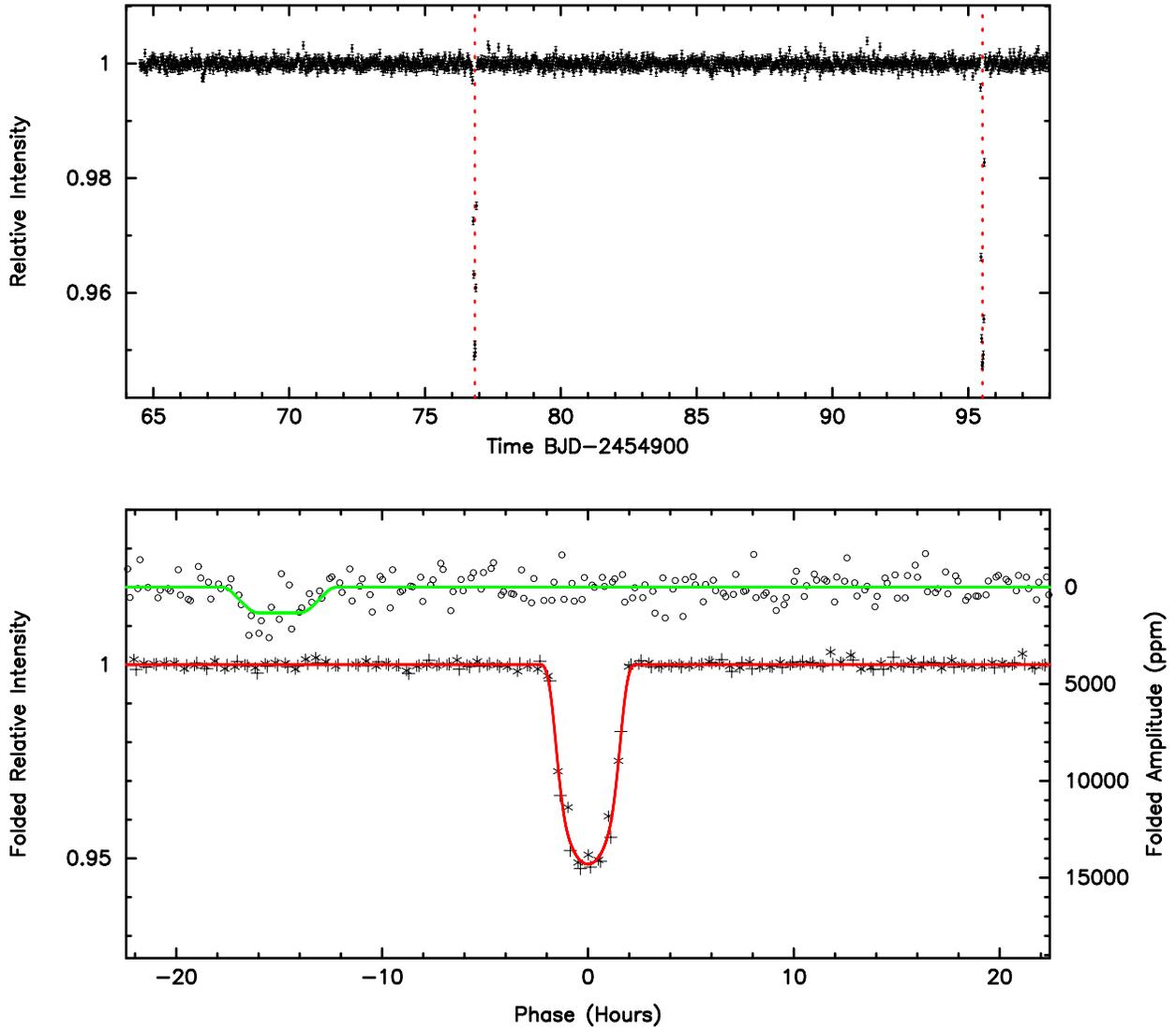}
\caption{{\it Kepler} light curve for KID-9838975. See Figure 2 caption. The scale for the occultation has been multiplied by a factor of 5 relative to the transit. The plot has two vertical axes scales, the left one is for the transit depth (black
crosses) while the right scale is for the secondary eclipse (grey circles). This object is an
eclipsing binary.}
\end{figure}

%Fig. 4
\begin{figure}
%\plotone{K_217.eps}
\plotone{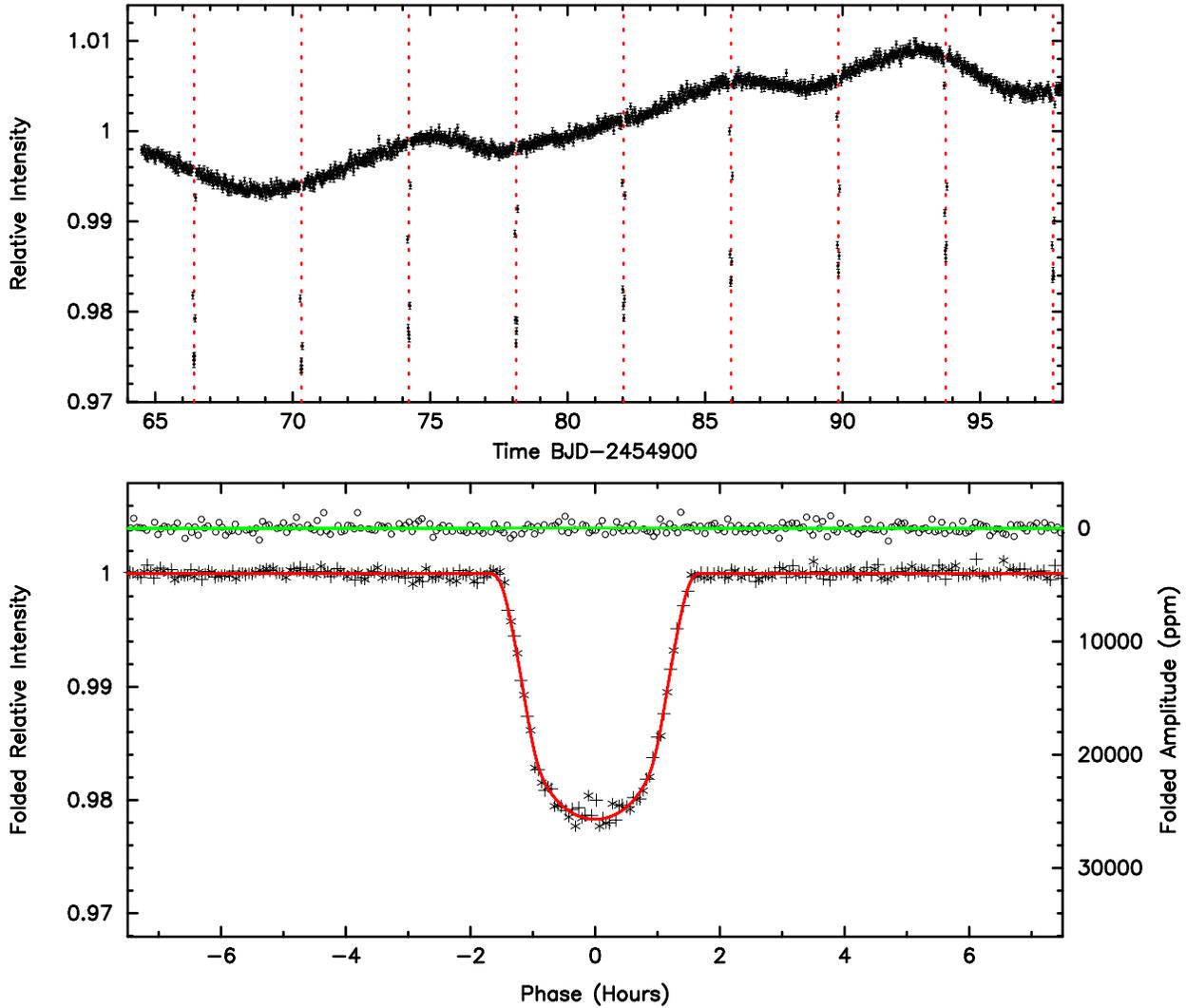}
\caption{ {\it Kepler} light curve for KID-9595827. See Figure 2 caption. Note the wiggles in the light curve at the bottom of the transit indicative
of star spots on the stellar surface. The plot has two vertical axes scales, the left one is for the transit depth (black
crosses) while the right scale is for the secondary eclipse (grey circles).
This object is a G8V star with a probable 1.12 Jupiter radius exoplanet 
in a 3.9 day orbit. }
\end{figure}

%Fig. 5
\begin{figure}
%\plotone{koi217.diff.ps}
\plotone{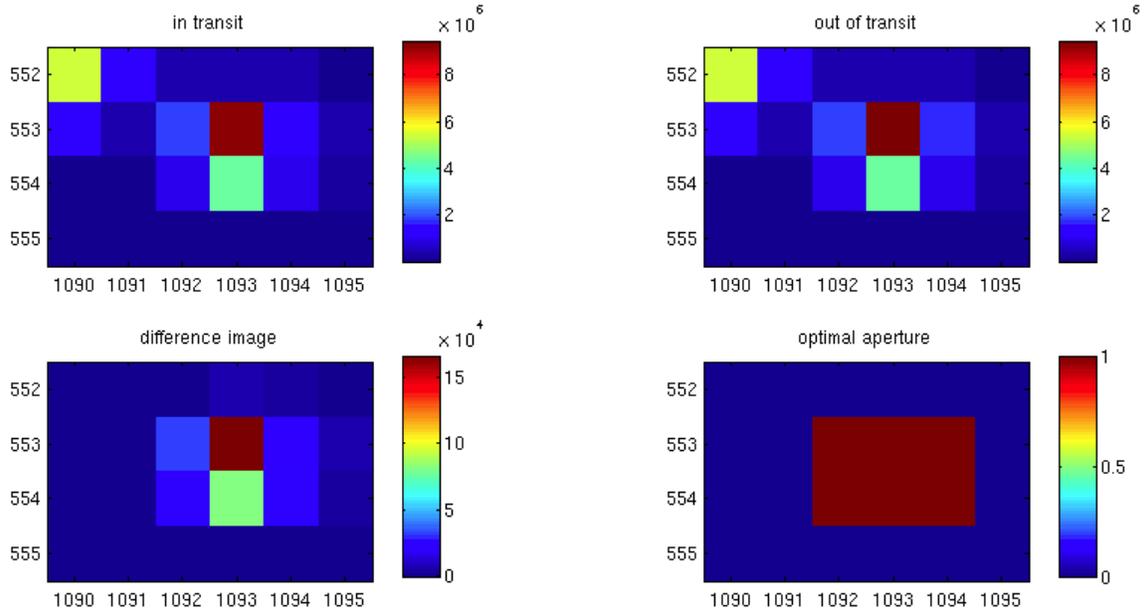}
\caption{{\it Kepler} pixel data images of KIC~9595827 (BOKS-1).
The top panels show an in and out of transit image while the bottom left panel
shows the difference image between the two. It is clear in the top two panels that 
a second star is (partially) present within the aperture for KIC~9595827 and 
its constant
light accounts for the small but measurable centroid shift observed during transit.
The difference image shows, however, that it is clearly only KIC~9595827 that varies
during the exoplanet transit.}
\end{figure}

%Fig. 6
\begin{figure}
%\plotone{pixellc.ps}
\plotone{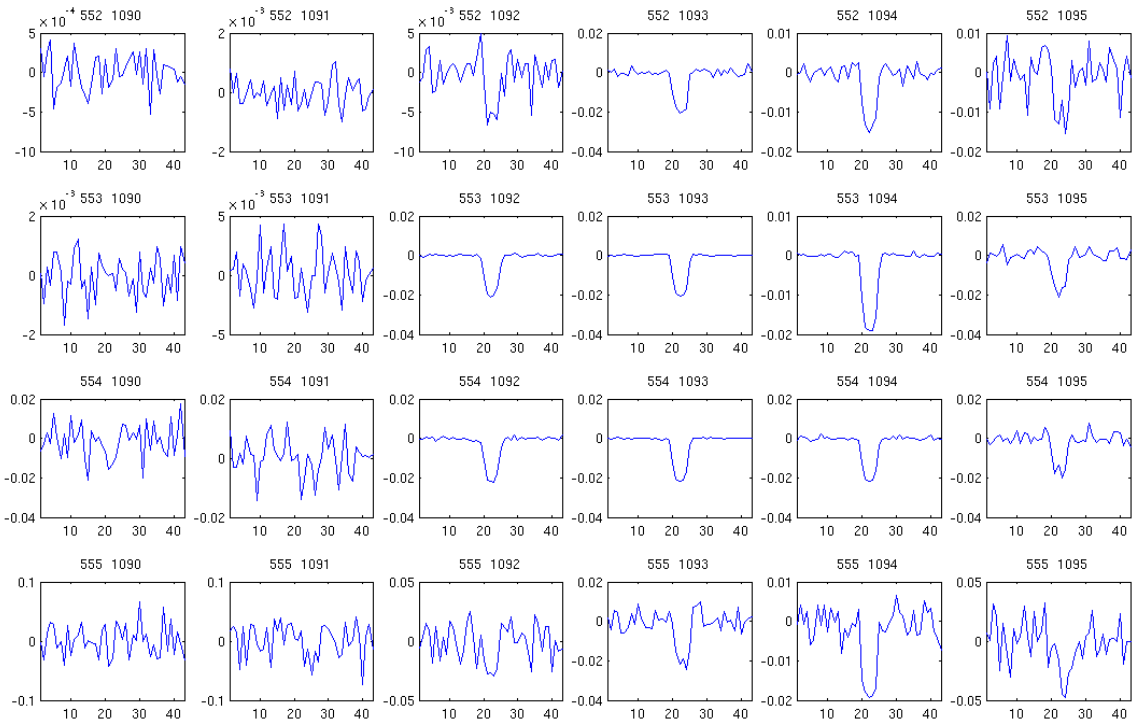}
\caption{Folded, median-normalized pixel time series for the {\it Kepler} aperture
containing KIC~9595827 and its neighbor. The pixel time series clearly shows which
pixels contain the transit signal. 
}
\end{figure}

%Fig. 7
\begin{figure}
%\plotone{K_SPplot.ps}
\plotone{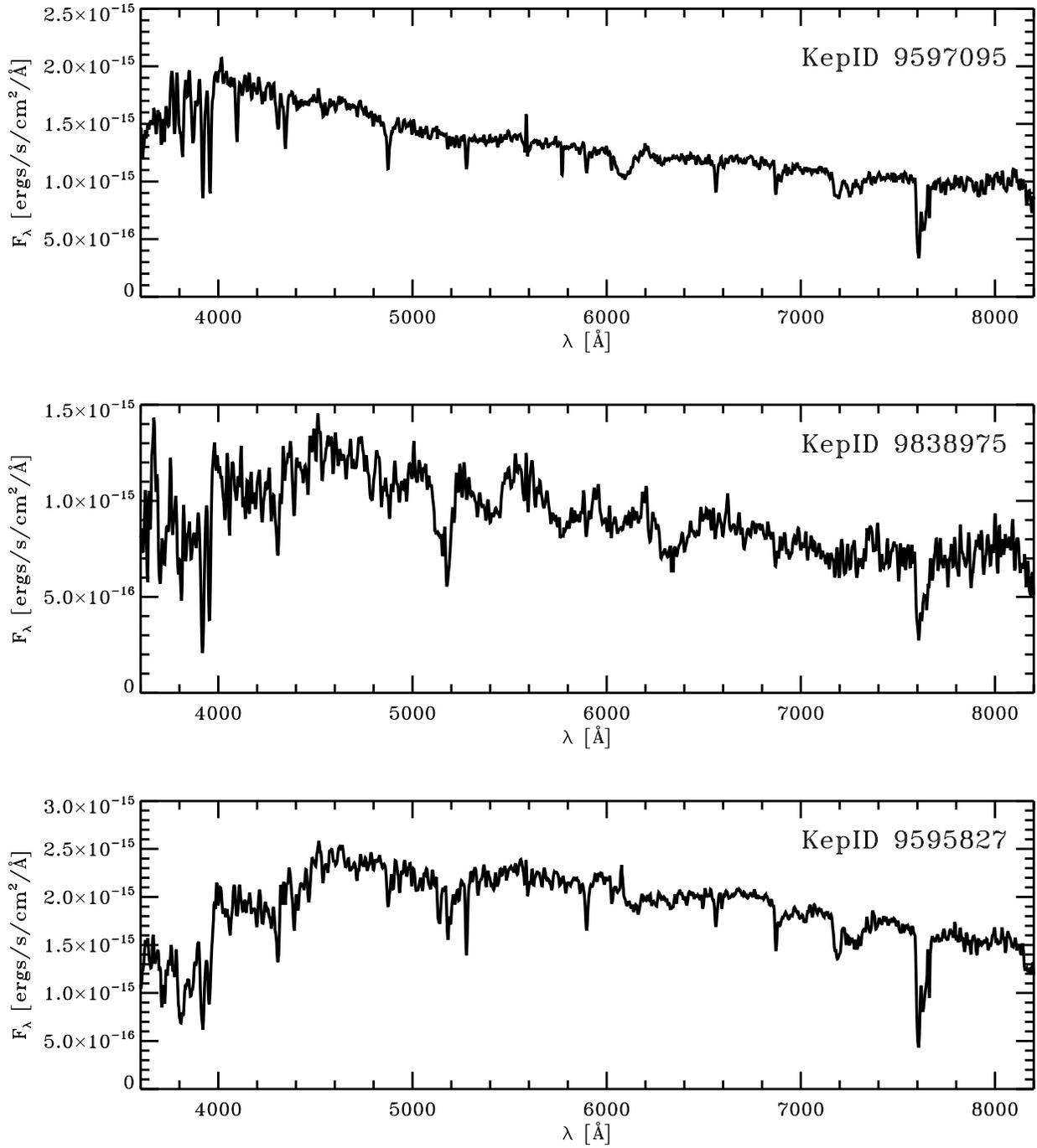}
\caption{Example Optical Spectroscopy of the three exoplanet candidate stars obtained at the Kitt
Peak 4-m telescope in June 2008.} 
\end{figure}

%Fig. 8
\begin{figure}
%\plotone{K_speckle1.ps}
\plotone{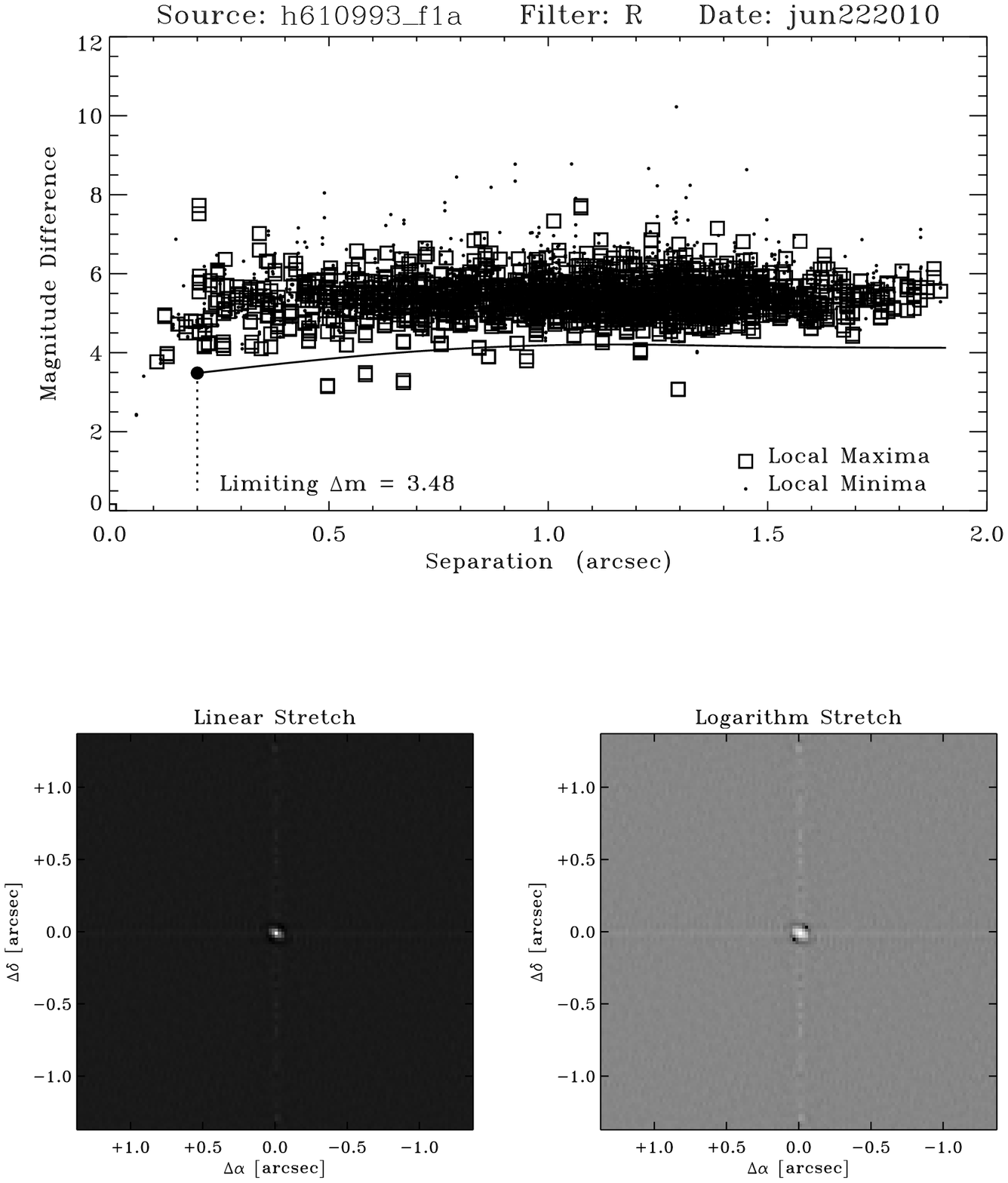}
\caption{ Speckle image of KID-9595827 showing that no line-of-sight or real companions exist
from 0.05 to 2.8 arcsec of the star to a limit of 4.2 magnitudes in R and 5.0 magnitudes in V fainter than the star itself.
The reconstructed images at the bottom of the plot have N up and E left. The
horizontal line in the top plot shows the 5 sigma detection limit for companions
against the sky background (with three reconstruction artifact points lying 
below) and the vertical line at 0.2 arc seconds is added to show the inner 
limit for conservative multi-fringe detections. See text for details.
}
\end{figure}

\end{document}